\documentclass[12pt]{article}
\usepackage{epsf, cite, amsmath, amssymb}
\usepackage{epsfig}

\makeatletter
\renewcommand\section{\@startsection {section}{1}{\z@}%
                                   {-3.5ex \@plus -1ex \@minus -.2ex}%nn
                                   {2.3ex \@plus.2ex}%
                                   {\normalfont\large\bfseries}}
\renewcommand\subsection{\@startsection{subsection}{2}{\z@}%
                                     {-3.25ex\@plus -1ex \@minus -.2ex}%
                                     {1.5ex \@plus .2ex}%
                                     {\normalfont\bfseries}}
\makeatother

\newcommand{\bea}{\begin{eqnarray}}
\newcommand{\eea}{\end{eqnarray}}
\newcommand{\be}{\begin{equation}}
\newcommand{\ee}{\end{equation}}

\newcommand{\T}{\theta}
\newcommand{\G}{\Gamma}
\newcommand{\F}{\Psi}
\newcommand{\e}{\epsilon}
\newcommand{\ce}{\overline{\epsilon}}
\newcommand{\LL}{\Lambda}
\newcommand{\dd}{\delta}
\newcommand{\com}[2]{{ \left[ #1, #2 \right] }} 
 
\newcommand{\rr}{\rightarrow}
\newcommand{\m}{\mu}
\newcommand{\n}{\nu}
\newcommand{\p}{\partial}
\newcommand{\hA}{\widehat{A}}
\newcommand{\hF}{\widehat{\F}}
\newcommand{\LT}{{\LL_\T}}
 
\newcommand{\C}[1]{$(\ref{#1})$}

\begin{document}
\begin{titlepage}

\begin{center}

{January 30, 2002}
\hfill                  hep-th/0201259

\hfill CGTP-02-01, EFI-02-61, UTTG-01-02

\vskip 2 cm
{\Large \bf I \, Non-commutativity and Supersymmetry}\\
\vskip 1.25 cm 
{
Sonia Paban$^1$, Savdeep Sethi$^2$ and Mark Stern$^3$}\\
\vskip 0.5cm
{\sl{}
$^1$ Department of Physics, University of Texas, Austin, 
TX 78712, USA \\
\vskip 0.2cm
$^2$ Enrico Fermi Institute, University of Chicago, Chicago, IL
60637, USA\\ 
and \\
School of Natural Sciences, Institute for Advanced Study, Princeton, NJ 08540
\vskip 0.2cm
{}$^3$ Department of Mathematics, Duke University, Durham, NC
27708, USA \\}

\end{center}

\vskip 2 cm

\begin{abstract}

\baselineskip=18pt

We study the extent to which the gauge symmetry of abelian Yang-Mills
can be deformed under two conditions: first, that the deformation depend
on a two-form scale. Second, that the deformation preserve supersymmetry.
We show that (up to a single parameter) the only allowed deformation is the one
determined by the star product. 
We then consider the supersymmetry algebra satisfied by NCYM expressed in 
commutative variables. The algebra is peculiar since the supercharges are
not gauge-invariant. However, the action, expressed in commutative variables,
appears to be quadratic in fermions to all orders in $ \T$.

\end{abstract}

\end{titlepage}

\pagestyle{plain}
%\baselineskip=18pt
% Try a wider skip
\baselineskip=20pt

\section{Introduction}

The appearance in string theory of non-commutative Yang-Mills~\cite{Connes:1998cr}\ leads us to a natural
question: how unique is the deformation that leads from commutative to non-commutative
Yang-Mills? It is natural to consider the question in the context of supersymmetric theories 
since they are most likely to exist at the quantum level. 
The cleanest example is N=1 Yang-Mills in ten dimensions. 
The goal of this work is to ask how we 
can deform this SYM (supersymmetric Yang-Mills) theory while still preserving
supersymmetry. In ten dimensions, the gauge coupling constant, $g^2$,
is dimensionful. However, we will not consider deformations which depend on higher powers 
of the coupling constant. Rather, we introduce a two-form scale $\T_{ij}$
with mass dimension $-2$ and consider deformations that depend on this scale.
For simplicity, we also restrict to deformations of abelian Yang-Mills.
Our goal is to understand to what extent the non-commutative generalization is determined 
by supersymmetry. For a review of earlier work on non-commutative field theory, 
see~\cite{Douglas:2001ba}. For work on superspace and non-commutative geometry, 
see~\cite{Klemm:2001yu}. 

In section two, we begin by recalling the structure of abelian SYM. We then proceed to describe the 
form and constraints on deformations compatible with supersymmetry. 
The deformations we permit are quite general, and can, in principle, look quite bizarre. 
{}For example, fermion bilinears can appear in the deformations. We also make no special 
assumptions about the structure of the deformation (like associativity of a product).
In section three, we undertake the
task of unraveling the constraints. To $O(\T)$, we show that the unique deformation of the gauge 
symmetry compatible with supersymmetry is precisely the deformation given by the star product 
commutator, 
\bea
\dd_\LL A_\m  &=& \LL_{,\m} + \T^{ij} \left( \p_i \LL \p_j A_\m - \p_i A_\m \p_j \LL  \right), \cr
&=& \LL_{,\m} + f_1(\LL, A_\m)
\eea
where $\dd_\LL$ generates a gauge transformation. This is a closed algebra at $O(\T)$ and no
terms with higher powers of $\T$ are required. There is also no free parameter that cannot be 
absorbed by rescaling $\T$. 

Beyond $O(\T)$, it makes the analysis considerably easier
to impose the condition that the supercharges, $\dd_\e$, be gauge-invariant,
\begin{equation}
[\dd_\LL, \dd_\e]=0,
\end{equation}   
as an off-shell condition. In both unperturbed abelian and non-abelian SYM theories, 
this condition indeed holds off-shell without use of the equations of motion. Under this
restriction, we show that by the use of field redefinitions, a gauge transformation 
can always be brought into the form
\begin{equation}
\dd_\LL A_\m  = \LL_{,\m} + f_1(\LL, A_\m) +  f_3(\LL, A_\m) + \sum_{p>3} f_p(\LL, A_\m),
\end{equation}   
where $f_i$ is an $O(\T^i)$ bilinear. An analogous statement holds for $\dd_\LL$ acting on a fermion. 
Each $f_i$ is completely determined, and agrees with the form
predicted from the star product commutator. The overall coefficient of each $f_i$ for $i>3$ is determined
relative to the coefficient of $f_3$. The only freedom is the overall 
coefficient of $f_3$. This is the only free parameter permitted in the deformation. 

We expect our argument to descend with little modification to lower dimensional theories.
In these theories (for example, N=4 D=4 SYM), it then appears that proving perturbative 
renormalizability reduces to showing that this one coefficient is invariant under renormalization
group flow. We should also point out that while supersymmetry is a key ingredient in our
argument, maximal supersymmetry is not. It would be interesting to see how few supercharges
are needed to guarantee the star product structure. Another direction worth mentioning is the
extension of these results to the effective action of N=4 SYM. Combining this analysis with
the techniques in \cite{Paban:1998ea}\ should lead to a non-perturbative way of 
obtaining gauge-invariant $F^4$ and $F^6$ interactions, extending 
the work of~\cite{Liu:2000ad, Liu:2000mj, Pernici:2000va}. 

In this analysis, we do not constrain higher derivative gauge invariant deformations
of the action. From the work of~\cite{Metsaev:by ,Metsaev:1987qp} on the DBI action
and supersymmetry, and the explicit supersymmetrization of DBI 
given in~\cite{Aganagic:1997nn}, we certainly expect constraints on higher derivative 
interactions.

In section four, we turn our attention to a puzzle. In~\cite{Seiberg:1999vs}, it is 
argued that the non-commutative theory can be rewritten in terms of commutative variables. 
{}From the commutative perspective, the action contains an infinite string of irrelevant
interactions. The structure of these interactions does not seem particularly remarkable. 
Yet if the deformation of the gauge symmetry
is so constrained by supersymmetry, we should ask: how do these constraints 
appear in terms of commutative variables? What we argue is that the usual supersymmetry
algebra is modified in a peculiar way. In particular, the supercharges, $\dd_\e$,  are no longer
gauge-invariant but rather
\begin{equation}
[\dd_\LL, \dd_\e] = \dd_{\LL'}.
\end{equation}
This is reminiscent of the structure that appears in a supergravity theory, although
supersymmetry still seems to be a global symmetry. The second peculiarity is that the
action need not contain couplings that are more than quadratic in the fermions, i.e., the 
simplest extension of the map of~\cite{Seiberg:1999vs} to fermions leads only to quadratic
couplings in the action.  This is quite
different from what we might have expected in a theory with an infinite number of higher 
derivative terms. The supersymmetry algebra is correspondingly highly constrained. 
To what
extent these structures generalize to a $D=6$ chiral tensor multiplet is a topic of current
investigation~\cite{toappear}. 

\section{N=1 D=10 SYM} 
\subsection{The undeformed abelian theory}

Let us begin by reviewing the structure of abelian N=1 SYM, and establishing our
notation. We choose a spacetime metric $\eta_{\mu\nu} =
(-,+,\ldots,+)$. Our only symmetry group is the Lorentz group
$SO(9,1)$.  
In terms of real gamma
matrices $\G_{\mu} \, , \mu=0,...,9$ satisfying, 
\begin{equation}\label{gamalg}
{ \{\G_{\mu},\G_{\nu}\}= 2\, \eta_{\mu \nu}, }
\end{equation}
we can express the supersymmetry variation of the gauge-field $A_\mu$ and 
real fermion $\F$:
\bea\label{transforms}
& &  \dd_\e A_{\mu}  = \frac{ i}{ 2}\,  \ce \, \G_{\mu} \F  \cr 
& & \dd_\e \F  =  - {1 \over 4} F_{\mu \nu} \G^{\mu \nu} \e. 
\eea 
We define $\ce = \e^{T} \G^0$ as usual, and the fermion is Weyl so there are
$16$ real components. The dimension of $A_\mu$ is $1$ while the dimension of 
$\F$ is $3/2$. 
Let us explicitly check closure of the supersymmetry algebra. The supersymmetry
algebra only closes up to gauge transformations, which we denote by $\dd_\LL$.
The algebra must satisfy,  
\bea\label{constraints}
& & [ \dd_{\LL}, \dd_{\e}] =0 \cr
& & [\dd_{\e_1},\dd_{\e_2}] = {i\over 2} \ce_2 \G^{\mu} \e_1  \partial_{\mu} + \dd_{\LL}.
\eea
The first condition is the statement that there is a well-defined global 
supercharge, independent of any gauge choice.  The second condition is simply the statement
that we have a closed symmetry algebra. Note, however, that closure is on Poincar\'e together
with a gauge transformation. This does not make geometric sense for non-commutative (or non-abelian) 
theories since we expect closure
on $D_\mu$ rather than $\partial_\mu$. As we shall see, the algebra does close on the geometrically
sensible covariant derivative. This occurs because the gauge parameter, $\LL$, appearing on the right hand
side of $(\ref{constraints})$ will actually become field-dependent. 
Lastly, we note that the gauge transformations themselves must leave invariant the equations of motion, and 
form a closed algebra.

On the gauge-fields, we see that
\bea\label{closureb}
 \com{\dd_{\e_1}}{\dd_{\e_2}} A_\mu & = &
 {i\over 2}  \ce_2 \G_{\nu}\e_1 \partial^{\nu}  A_{\mu} 
-  {i\over 2}  \partial_{\mu} ( \ce_2 \G_{\nu} A^{\nu} \e_1) , \cr
& = & {i\over 2}  \ce_2 \G^{\nu}\e_1 F_{\nu\mu}. \eea
For this to satisfy $(\ref{constraints})$, we see that
\begin{equation}\label{ungauge}
{\dd_{\LL} A_{\mu} =  -{i\over 2}\partial _{\mu} (  \ce_2 \G_{\nu} A^{\nu} \e_1)}
\end{equation}
for the undeformed theory. This is a conventional gauge transformation but the gauge 
parameter is field-dependent as promised. Note that the right hand side of $(\ref{closureb})$ 
depends only on the gauge-invariant combination, $F_{\mu\nu}$. 
This is worth noting because when we come to the five-brane, the correct gauge-invariant
objects to study are unknown.

Closure of the supersymmetry algebra on $\F$ gives,

\bea\label{closureF}
\com{\dd_{\e_1}}{\dd_{\e_2}} \F & = {i\over 2} \ce_1
 \G^{\mu} \e_2 \partial_{\mu} \F + \ldots. \eea
% - i{ 5 \over 64} \ce_1 \G^{\mu} \e_2  \G_{\mu
%\nu} \partial^{\nu} \f \cr & -  {i \over 1536} \ce_1  \G^{\mu \nu \alpha \beta \gamma}
%\e_2  \G_{\mu \nu \alpha \beta } \partial_{\gamma} \f  }}
The omitted terms are all proportional to the equation of motion for
$\F$ so, as usual, the algebra only closes on-shell.

\subsection{The deformed abelian theory}

How can we deform the supersymmetry transformations? The most general 
deformation can be captured by the following modifications, 
\bea\label{deformsusy} & & \dd_\e A_{\mu}  ={ i \over 2}\,  \ce \, \G_{\mu} \F + \ce 
\, N_{\mu} \F \cr 
& & \dd_\e \F  = - {1 \over 4} F_{\mu \nu} \G^{\mu \nu} \e + M \e,
\eea
where $N$ and $M$ are unspecified operators. It will often be convenient to rewrite 
these transformations without explicit epsilons, 
 \bea\label{deform} & & \dd_a A_{\mu}  ={ i \over 2}\,  (\G^0 \G_{\mu} \F)_a +  
 (\G^0 N_{\mu} \F)_a \cr 
& & \dd_a \F_b  = - {1 \over 4} F_{\mu \nu} \G^{\mu \nu}_{ba} + M_{ba},
\eea
We do not want to expand $N$ and $M$ in derivatives, but we can classify
terms in $N$ or $M$ by number of fermions. Each of these terms can have an 
arbitrary number of derivatives. The natural physical constraint is that $N, M$
vanish when $\T_{ij} \rr 0$. 

Field redefinitions play a crucial role in our analysis. For $N_\m$ of the form 
$$N_\m \sim \G_\m ({\rm anything}),$$ 
where $($anything$)$ is Lorentz invariant, we can make a field redefinition
$$ \widetilde{\F} = \left(1+{\rm anything}\right)\F $$
to remove this kind of $N_\m$. Unfortunately, not all $N_\m$ are of
this convenient form. We also need to keep in mind possible field redefinitions
of the gauge-field, 
$$ \widetilde{A_\m} = A_\m + h_\m (\T, A, \F),$$
where $h_\m \rightarrow 0$ as $\T\rightarrow 0$, and $h_\m$ is gauge invariant. 
{}For recent comments on controlling
field redefinitions in supersymmetric theories, see~\cite{Cederwall:2001dx}.  

In studying non-trivial deformations, we must also consider the possibility that
our gauge symmetry is modified.   Let us
express our modified gauge symmetry in the following way, 
\bea\label{modgauge}
& &\delta_{\LL} \, A_\mu =  \LL_{\, ,\mu} + B^{\LL}_{\mu}, \cr
& &\delta_{\LL} \, \F_a  = C^{\LL}_a, 
\eea
where $B$ and $C$ are at least of order $\T$. Where convenient, we will express 
the modified gauge transformation in a power series in $\T$, 
$$ \dd_\LL =  \dd_\LL^0 + \dd_\LL^{1} + \dd_\LL^{2} + \ldots, $$
where $ \dd_\LL^{n}$ is $O(\T^n)$.  Also note that, 
\begin{equation}\label{gaugecomm}  [ \dd_{\LL}, \dd_{\LL'}] = \dd_{\LL_\T}, \end{equation}
where $\LL_\T$ is at least 
$O(\T)$ because we started with an abelian theory so the $O(0)$ terms commute. 

\subsection{The general form of the closure constraints}

We start with the relation, 
\begin{equation}\label{linear}  
\delta_{\LL_1}  +\delta_{\LL_2}  = \delta_{\LL_1+\LL_2}. \end{equation}
This is forced by considering the lowest unperturbed gauge transformations. However, this
constraint implies that $\delta_\LL$ is linear in the gauge parameter $\LL$. We still require
that the supercharges be well-defined with respect to the gauge symmetry so, 
\begin{equation}\label{bigdeal} 
[ \dd_{\LL}, \dd_{a}] =0. \end{equation}
In this, and all subsequent discussion, we must keep in mind that the ``$0$'' appearing on the
right hand side could include terms that vanish on-shell. 
If \C{bigdeal} is not true then the supercharges are not gauge-invariant. In most cases, this implies that 
supersymmetry is part of the gauge symmetry, and we are forced to 
consider a theory of supergravity. We will therefore impose this condition. 
Let us apply this constraint first to the gauge-field, 
\bea
 [ \dd_{\LL}, \dd_{a}] A_\mu & = &  \dd_{\LL} (\G^0 \left\{ {i\over 2}\G_\mu \F + N_\m \F \right\} )_a 
- \dd_a B^\LL_\mu, \cr
& = & {i\over 2} (\G^0 \G_\mu C^\LL )_a + \dd_{\LL} (\G^0 N_\m \F)_a 
- \dd_a B^\LL_\mu, \cr
& = & 0.
\eea
This gives us the following relation for $C$ in terms of $B$: 
\begin{equation}\label{solveC} 
C_a^\LL = 2i (\G^\mu \G^0\, \dd)_a B^\LL_\mu - 2i \, \dd_{\LL} (\G^\m N_\m \F)_a . \end{equation}
Note that we do {\it not} have to sum on $\mu$ in $(\ref{solveC})$. However, the expression for $C^\LL$ must 
be independent of $\mu.$

Applying \C{bigdeal} to the fermion gives the following relation, 
\bea\label{ferminv} 
[ \dd_{\LL}, \dd_{a}] \F_b & = & \dd_\LL ( -{1\over 4} F_{\mu \nu} \G^{\mu \nu} + M )_{ba} - \dd_a C^\LL_b, \cr
& =& 0. \eea

We now turn to closure of the gauge algebra. Acting on the fermion $\F$, we obtain the following relations
\bea\label{cclosure} 
[ \dd_{\LL'}, \dd_{\LL}] \F_a &  = &  \dd_{\LL'} C^{\LL}_a -   \dd_{\LL} C^{\LL'}_a, \cr
&  =&  {\dd C^\LL_a \over \dd A_\n} ( \LL'_{,\n} + B^{\LL'}_\n) - {\dd
C^{\LL'}_a \over \dd A_\n} ( \LL_{,\n} + B^{\LL}_\n) +
{\dd C^\LL_a \over \dd \F_b} ( C^{\LL'}_b ) - {\dd C^{\LL'}_a \over \dd \F_b} ( C^{\LL}_b ), \cr
&  =& \dd_{\LL_\T} \F = C^{\LL_\T}_a. 
\eea
Note that the functional derivatives  ${\dd C^\LL_a \over \dd A_\n}(\ldots) $ and 
${\dd C^\LL_a \over \dd \F_b}(\ldots)$
are operators that act on their respective arguments.
Dimensional analysis is kind to us here. The term $ C^{\LL_\T}_a$ is
at least $O(\T^2)$. In a $\T$ expansion, the dominant term in  
$ C^{\LL_\T}_a$ is, 
\begin{equation}\label{domC} 
{\dd C^\LL_a \over \dd A_\n} ( \LL'_{,\n}) - {\dd C^{\LL'}_a \over \dd A_\n} ( \LL_{,\n} ).
\end{equation}
This term is only $O(\T)$ and must vanish.

Closure of the gauge algebra on the gauge-fields gives us the relation, 
\bea\label{Aclosure} 
[ \dd_{\LL'}, \dd_{\LL}] A_\m &  = &  \dd_{\LL'} B^{\LL}_\m -   \dd_{\LL} B^{\LL'}_\m, \cr
&  =&  {\dd B^\LL_\m \over \dd A_\n} ( \LL'_{,\n} + B^{\LL'}_\n) - {\dd
B^{\LL'}_\m \over \dd A_\n} ( \LL_{,\n} + B^{\LL}_\n) +
{\dd B^\LL_\m \over \dd \F_b} ( C^{\LL'}_b ) - {\dd B^{\LL'}_\m \over \dd \F_b} ( C^{\LL}_b ), \cr
&  =& \dd_{\LL_\T} A_\m = \LL_{\T,\m} +  B^{\LL_\T}_\m. 
\eea
There are further conditions from closure of the supersymmetry algebra which we shall describe
when needed. 

\section{Unraveling the Constraints}
\subsection{The gauge parameter $\LL_\T$}

Our first step is to learn something about the gauge parameter $\LL_\T$ that appears in the
commutator, 
$$ \com{\dd_\LL}{\dd_{\LL'}} = \dd_{\LL_\T}, $$
where $\LL, \LL'$ are field-independent. We know that $\LT$ is an antisymmetric bilinear form
in  $\LL$ and $\LL'$. 
What we wish to show is that $\LL_\T$ is 
field-independent. Using \C{bigdeal}, it is easy to show that
\begin{equation}\label{susycon} \com{\dd_a}{\dd_{\LL_\T}} = 0, \end{equation}
up to terms that vanish on-shell. The on-shell caveat is needed because the SUSY algebra typically 
only closes on-shell. The parameter which determines the deformation of our gauge symmetry 
therefore commutes with the SUSY vector-fields, $\dd_a$. 

If we apply \C{susycon} to $A_\m$, we learn that 
\bea  \com{\dd_a}{\dd_{\LL_\T}} A_\m  &= & \dd_a (\LT_{,\m} + B_\m^{\LL_\T}) 
- \dd_\LT ({ i \over 2}\,  
(\G^0 \G_{\mu} \F)_a +  (\G^0 N_{\mu} \F)_a), \cr
& = & \dd_a^0 (\LT_{,\m}) + O(\T^2), \cr &=& 0.
\label{paramconst}\eea
At $O(\T)$, there is no freedom to construct a term in $\LT$ 
from fermions on dimension grounds. Any term 
constructed from bosons must vanish by \C{paramconst}. This statement is true on the nose rather than
just on-shell: the
only term that could vary into an equation of motion has the form,
$$ \p^\n A_\n \T \LL \LL'. $$
This is neither antisymmetric in $\LL, \LL'$ nor is there a way to contract the indices. 
So $\LT$ is field independent to $O(\T)$.

To go beyond $O(\T)$, we note that \C{paramconst} is satisfied for any 
$\LT$ independent of $A, \F$. This teaches us that, 
$$ \dd_a (\LT_{,\m}) + B_\m^{\dd_a \LT} = 0. $$
Now this is a trianglar set of equations which 
reduce to the statement
\begin{equation}\label{smalleq} \dd_a^0  (\LT)= 0.\end{equation}
If we apply a second $\dd_a^0$ and use closure of the supersymmetry algebra at 
lowest order, we see that 
$$ \dd_a^0 \dd_a^0 (\LT) \sim (\G^\n_{aa} \p_\n + \dd_{\G^\n_{aa} A_\n}^0) \LT =0.$$
We can then conclude that $\LT$ is independent of fermions (on which $\dd^0_\LL$ 
vanishes). Returning to \C{smalleq} then teaches us that $\LT$ is independent of
bosons. We therefore conclude that $\LT$ 
is field independent to all orders in $\T$ up to terms whose variation is zero on-shell. 
We now impose \C{bigdeal} as an off-shell constraint. Under this stronger constraint, 
$\LT$ must be field independent to all orders in $\T$.   

\subsection{The Jacobi identity}

Now that we have established that $\LT$ is field independent, we can try to construct 
$\LT$ in terms of $\LL,\LL', \T$ and derivatives. Let us write the gauge parameter $\LT$
in terms of a function $f$, 
\begin{equation}
\LT = f(\LL, \LL'),
\end{equation}
where $f$ is antisymmetric in its two arguments. To order $\T$, it is easy to see that 
there is a unique possibility,
\begin{equation}
f(\LL, \LL') = \T^{ij} (\p_i \LL \p_j \LL' - \p_i \LL' \p_j \LL) + O(\T^2). 
\end{equation} 
This leading term itself defines a closed supersymmetry algebra. Nevertheless, we can ask 
about the structure of possible higher $\T$ corrections to $f$.    

In performing this analysis, it is rather crucial that the Jacobi identity is obeyed by 
$f$. Since we deal with standard 
operator composition, our gauge transformations automatically satisfy the usual Jacobi
identity
$$ [\dd_{\LL}, [\dd_{\LL'}, \dd_{\LL''}]] + {\rm cyclic} = 0.$$
In turn, this relation implies that
\begin{equation} \label{jacobi}
f(\LL, f(\LL', \LL'')) + f(\LL', f(\LL'', \LL)) + f(\LL'', f(\LL, \LL')) =0.
\end{equation}

\subsection{Solving for $\LT$.}

At order $\T$, we introduce no free parameters. Any choice can be
absorbed into a redefinition of $\T$. Beyond $O(\T)$, we will
introduce parameters. The key question in understanding the 
renormalizability and uniqueness of our deformation is, how many?
An explicit computation shows that there are no $O(\T^2)$ deformations
of the gauge algebra consistent with \C{jacobi}. The first non-trivial 
deformation is $O(\T^3)$. 

In order to analyze the structure of allowed deformations, we expand $f$ in a 
series,   
\begin{equation}\label{series}
f = f_1 + f_3 + \sum_{p>3}f_p, 
\end{equation}
where each $f_k$ is homogeneous of order $k$ in $\T$.  
Since $\T$ has mass dimension $-2$, each $f_k$ must be homogeneous of 
degree $2k$ in derivatives. Since $f$ is antisymmetric, so too 
is each $f_k$. Since we will consider only antisymmetric functions,  
it is convenient to adopt exterior algebra notation. 
We therefore write,  
$$f_k = m_k^{IJ}\p_I\wedge\p_J.$$
Here $I=\{i_1,\cdots,i_n\}$ and $J$ are multi-indices. We define 
$$|I| = i_1+\cdots + i_n, \quad \p_I = \p_1^{i_1}\cdots\p_n^{i_n},$$ 
and $m_k^{IJ}$ is zero unless $|I|+|J| = 2k$. When $I$ is empty, we set
 $\p_I = 1$. 
The exterior algebra notation is to be interpreted as usual, 
$$f_k(x,y) = m_k^{IJ}\p_I(x)\wedge\p_J(y) - m_k^{IJ}\p_I(y)\wedge\p_J(x).$$
This allows us to view differential operators as differential forms. 
We will similarly define $3$-form differential operators 
\bea 
\p_I\wedge\p_J\wedge\p_K(x,y,z)  =  & & \p_I(x)\wedge\p_J(y)\wedge\p_K(z)
+ \p_I(y)\wedge\p_J(z)\wedge\p_K(x) + \cr 
& &  \p_I(z)\wedge\p_J(x)\wedge\p_K(y). 
\eea
The generalization to higher forms is straightforward. 

We can now express the Jacobi identity in terms of our expanded $f$. 
It gives the equalities 
\begin{equation}
\label{exjac}
\sum_{i+j=k+1}(f_i(f_j(x,y),z) + f_i(f_j(y,z),x) + f_i(f_j(z,x),y))
 = 0. \end{equation}
We can view this equation as inductively defining 
$f_k$ in terms of the $f_j,$ $j<k$.  To what extent are the $f_j$ uniquely
determined? We started with the abelian Lie algebra of functions. 
Turning on $f_1$ deforms this to the Poisson Lie algebra, 
which has $1$ free parameter associated with scaling the symplectic form $\T$. 
For the Lie algebra determined by the $*$-product, this is the only free 
parameter. 

However, since the $O(\T)$ deformation gives a Lie algebra without need for any 
higher terms, we also have a free parameter which corresponds to scaling $f_3$. 
What we will show is that there are no other 
free parameters. If the coefficient of the $f_3$ deformation agrees with the 
value predicted by the $*$-product then our deformation {\it must} lead
to non-commutative Yang-Mills. 

In order to prove that all $f_k$ for $k>3$ are determined uniquely, 
it suffices to prove the stronger statement that there exists at most one 
$f_k$ satisfying \C{exjac}. 
%We remark that this also gives a simple proof of 
%the Lie algebra version of Kontsevich's theorem on the uniqueness of 
%the star product \rkont.
If given $f_j$ for $j<k$, there exist two distinct functions which can be 
substituted for $f_k$ in \C{exjac}, then taking their difference gives a 
function $g$ satisfying 
\begin{equation}
\label{exjacg}
  f_1(g(x,y),z) + g(f_1(x,y),z)  + {\rm cyclic}  = 0.
\end{equation}
This equation is satisfied by $g=f_1$ and $g=f_3$. We will show it has no
other solutions. 
As before, we write 
$$g = m^{IJ}\p_I\wedge \p_J$$
where $|I|+|J|=2k$ if $m^{IJ}$ is nonzero, and $m$ is antisymmetric.
We introduce one more notation. We write the Leibniz rule as follows: 
$$\p_I(BC) = l_I^{JK}\p_J(B)\p_K(C),$$
where $l_I^{JK}$ is defined by the Leibniz rule.
 
With this notation, we expand \C{exjacg}:   
\bea  
& & \theta^{ij}m^{IJ}[\p_i(\p_I\wedge \p_J)(x,y)\p_j(z) + \p_I(\p_i(x)\p_j(y))\p_J(z) - 
\p_I(\p_j(x)\p_i(y))\p_J(z) \cr & & +  \, {\rm cyclic}] = 0, \cr
& & \Rightarrow \quad 
\theta^{ij}m^{IJ}[\p_i\p_I\wedge \p_J\wedge\p_j + \p_I\wedge\p_i\p_J\wedge\p_j
+ l^{ab}_I\p_a\p_i\wedge \p_b\p_j\wedge P_J] = 0.
\eea
Now observe that there are many ways of interpreting this ``three-form.''
By evaluating at the origin, it can be construed as a genuine three-form 
acting on the vector space of polynomials. This allows 
us to use standard exterior algebra rules for evaluating. 

We note that the first $2$ terms cancel with the summands in the third term
where $|a|$ or $|b| = 0$. This leaves us with 
\begin{equation}\label{thepoint}
\theta^{ij}m^{IJ}\sum_{|a|,|b|>0} l^{ab}_I\p_a\p_i\wedge\p_b\p_j\wedge\p_J = 0.
\end{equation}
Let us skew diagonalize $\theta$,  and suppose that $\T^{12}$ is nonzero. 
Fix a degree $d<2k$, and consider a $\p_I$ of degree $|I|=d$ which appears
with nonzero $m^{IJ}$ coefficient, some $I$ 
 and the largest number of $\p_1$ factors. Call this $I:=I_t$. 
 If $d<|I_t|$, then we see that for $l>1$ there is no term to cancel 
$$\theta^{12}m^{I_t,J}\p_{I_t-\{l\}}\p_1\wedge \p_l\p_2\wedge \p_J$$
unless $|J|=2$.  So, if there exists no $|J|=2$ factor, $|I_t| = d$.
So, we first show that no such terms appear. 
If there is a degree two factor, say $\p_J=\p_s\p_t$, then we will get terms 
of the form 
$$\theta^{12}m^{JI}r^{st}_J\p_s\p_1\wedge\p_t\p_2\wedge\p_I = 0.$$

Once again consider the term $|J|=2$ with $j_1$ maximal. 
Arguing as before (now $|I|>2$ for $k>2$), $j_1=2$. So, the term 
$$2\theta^{12}m^{JI}\p_1^2\wedge\p_1\p_2\wedge\p_I = 0$$
must be cancelled, but obviously cannot be, so there are no 
$|J|=2$ terms.

Returning to our argument for general $|I|>2$,
 we obtained a contradiction unless
 $$|I_t|=d.$$
We now consider the contribution of the terms $m^{I_tJ}$ to \C{thepoint}.
These give terms, 
$$\theta^{12}m^{I_t,J}d\p_1^d\wedge \p_1\p_2\wedge \p_J.$$ 
It is easy to see that these terms cannot be cancelled by any term 
with $\p_I = \p_1^{d-1}\p_k$ for that would lead to a $\p_1^d$ term
of the form 
$$\theta^{12}m^{I,J}(d-1)\p_1^d\wedge \p_k\p_2\wedge \p_J,$$
which could not cancel for $k\not = 1$. 
Note the factors of $d$, $d-1$ above are just those determined by the Leibniz
rule.

Therefore, the only possible way to cancel this term is  by 
the contribution from some $m^{K,I_t}$ term. Such a term enters as
$$\theta^{ij}m^{K,I_t}\p_{K-\{l\}}\p_i\wedge \p_l\p_j\wedge \p_1^d.$$
{}For this to cancel the term in question, 
we need $K- ({l}\cup {i}) = J$ and $\p_l\p_j = \p_1\p_2.$ 
This implies $K=J$,  and gives us two terms which sum to give 
$$\theta^{12}m^{J,I_t}(j_1-j_2)\p_J\p_1\wedge \p_1\p_2\wedge \p_1^d,$$
where again $j_i$ is the exponent of $\p_i$ in $\p_J$.  
    
Hence, for cancellation, $j_2-j_1 = d.$ Suppose we had chosen $d\geq k$. 
Then we see from this equality that in fact $d=k$ and $j_2=k$, $j_1=0$. 
We therefore deduce that 
 $m^{IJ}=0$ unless $|I|=|J|=k$. 
To simplify further, we must go to the next term in the Leibniz rule expansion.
Let $J_t$ be given by $\p_{J_{t}} = \p_2^k$.
Consider the term 
$$\theta^{12}m^{I_t,J_t}\p_1^{k-1}\wedge \p_1^2\p_2\wedge \p_2^k.$$
As we have shown, only terms with $|I|=|J|$ enter. When $k>3$, 
this can only be cancelled against terms $J=J_t$, and
$\p_I = \p_1^{k-2}\p_t\p_s.$ In order to match the term $\p_1^2\p_2$ 
we must have $s=t=1$. So, there is no new term to cancel against 
the contribution from $m^{I_tJ_t}$. This leads to a contradiction 
unless $g$ is identically zero. Finally, we can conclude that the $f_k$ 
are unique up to a scaling of $f_3$. 
Since we know that the $*$-product gives one solution for the $f_k$ with $k\geq 3$, 
it must be the solution determined by supersymmetry.

\subsection{Determining $B$ and $C$}
\label{detbc}
Now that we have shown that the gauge parameter is the one determined by the star product
up to a constant, we need to ask whether $B_\m^\LL$ and $C_a^\LL$ can always
be made to agree with the expected forms,
\bea
(B_\m^\LL)_{ex}   & = & f(\LL,  A_\m), \cr
(C_a^\LL)_{ex} &= & f( \LL,  \F_a).
\eea
With a specific choice of scale for $f_3$,  
$$ f(A,B) = A\ast B - B\ast A, $$
where $\ast$ denotes the usual star product. 

Let us write $B$ and $C$ as perturbations of $B_{ex}$ and $C_{ex}$,
\bea
B^k &=& B_{ex}^k + b^k, \cr
C^k &=& C_{ex}^k + c^k, 
\eea
where the superscript, $k$, denotes terms of $O(\T^k)$. We will inductively show that
we can choose $b^k=0$ and $c^k=0$. So let us assume that for $j<k$, 
$$b^j=c^j=0.$$   
Now closure of the gauge algebra gives the relation, 
\bea
[\dd_\LL, \dd_{\LL'}] A_\m & = & \dd_\LL B_\m^{\LL'} - \dd_{\LL'} B_\m^{\LL}, \cr
& = & \dd_\LT A_\m =  \LT_{,\m} + B^\LT_\m. 
\eea
At $O(\T^k)$, this gauge closure relation and its analogue for $\F$ yield the conditions:
\bea \label{nconstraints}
& \dd_\LL^0 (b^{\LL'}_\m)^k - \dd_{\LL'}^0 (b^{\LL}_\m)^k & = 0, \cr
& \dd_\LL^0 (c^{\LL'}_a)^k - \dd_{\LL'}^0 (c^{\LL}_a)^k & = 0.
\eea
What we will show is that \C{nconstraints} implies that we can make field redefinitions 
that eliminate $b^k$ and $c^k$, with the possible introduction of new $b^l, c^l$ 
with $l>k.$ The idea is \C{nconstraints} should be viewed as saying that  
$b^k$ and $c^k$ are closed one-forms. We then seek an 
algebraic Poincar\'e lemma which says that they are in fact exact. By exact, we mean in 
the sense that there exists $\beta^k_\m, h^k_a$ so that 
$$\delta_\LL^0 \beta^k  = (b^\LL)^k,$$
and 
$$\delta_\LL^0 h^k  = (c^\LL)^k. $$
Then by replacing $A_\m$ by,
$$A_\m \rightarrow A_\m - \beta^k_\m,$$ 
and replacing $\F_a$ by, 
$$ \F_a \rightarrow \F_a - h^k_a, $$ 
we remove $b^k$ and $c^k$ from our gauge transformation.
 
Mimicking the usual Poincar\'e lemma, we will construct $\beta^k_\m$ and $h^k_a$ by 
``integrating'' $ (b^\LL)^k$ and $(c^\LL)^k$. We  replace derivatives of $\LL$ with 
appropriate terms in $A$. Terms with undifferentiated $\LL$ cannot be
integrated in this manner and so will have to be handled separately. 
The procedure to eliminate the terms with $\LL$ derivatives will be the same for $c^k$ 
and $b^k$ so we treat only the former. We further expand $c^k$  
$$c^k = \sum_pc^{kp}$$ 
where $c^{kp}$ is homogeneous of degree $p$ in $A$ and its derivatives. 

Equation \C{nconstraints} implies that 
we can rewrite $\delta_{\LL_1} (c^{\LL_2})^{kp}$ in the form, 
$$ \left\{ (\LL_1)_{,I} (\LL_2)_{,J} + (\LL_1)_{,J} (\LL_2)_{,I} \right\} c^{kpIJ}.$$ 
Note that $I,J,\ldots$ are spacetime indices. 
More generally, the fact that the
unperturbed gauge transformation is abelian implies that 
$$ \delta_{\LL_1} \cdots \delta_{\LL_{d-1} } (c^{\LL_d})^{kp} 
= (\LL_1)_{,I_1} \cdots (\LL_d)_{,I_d} c^{kpI_1 \cdots I_d},$$
with $c^{kpI_1\cdots I_d}$ totally symmetric in the $I_j$. 
Since $p$ is finite, there exists a maximal $d$ -- call it the depth --  so that 
$$ \delta_{\LL_1}\cdots\delta_{\LL_{d-1}} (c^{\LL_d})^{kp}$$ 
is nonzero, but all higher gauge variations vanish. In particular, 
for this $d$, $c^{kpI_1\cdots I_d}$ is gauge invariant. 

Let us define
$$h^{kpd} = {1\over d!} A_{I_1}\cdots A_{I_d}c^{kpI_1\cdots I_d}.$$
Then it is easy to see that making the change of variables 
$$ \F_a \rightarrow \F_a - h^{kpd}_a,$$
has the effect of replacing $c^k$ with a new $c^k$ whose maximal depth is less
than $d$. Performing induction on the depth, we can eventually reduce to the case 
$d=0$ which is the same as $c^k = 0$ -- at least for all terms in $c^k$ that 
depend on derivatives of the gauge parameter. A similar argument can be applied to 
$b^k$. 

We are left with the task of eliminating terms in $(c^\LL)^k$ which are of the form 
$ \LL c_0^k$.  Namely, terms that do not depend on derivatives of $\LL$. 
We return to the condition that the supercharges be gauge invariant,
$$[\delta_\LL,\delta_a] A_\m = 0,$$ 
which implies at $O(\theta^k)$ that
$$\left( 
\delta_{\LL}^j \delta_a^{k-j}  - \delta_{a}^j\delta_{\LL}^{k-j} \right) A_\m = 0.$$
Assume, inductively, that $k$ is the first order at which the gauge parameters
enter undifferentiated. Then the only terms with undifferentiated $\LL$ in the
preceding equality are 
$$ \LL (c_{0}^k)_t (\G_\m)_{ta} - \delta_{a}^0 \LL (b^k_{0})_\m = 0.$$
Thus, if we can show that $c_0^k = 0,$ we are left with the condition that 
$b_0^k$ is a susy invariant and therefore zero. Similarly, the vanishing of 
$(b^k_{0})_\m$ implies the vanishing of $c^k_0$. 

{}From the rigidity of the supercharges (that they are gauge-invariant) and 
from closure of the gauge algebra, we get 
\begin{equation} 
0 = [\delta_\LL,\{\delta_a,\delta_b\}] = 
[\delta_\LL,\G^\n_{ab}\p_\n + \delta_{\G^\n_{ab}A_\n} + \delta_v] = 
[\delta_\LL, \delta_{\G^\n_{ab}A_\n} + \delta_v].
\end{equation}
Here, $\dd_v$ contains all the corrections to the gauge transformation appearing in the
closure relation. We expand this relation out (acting on $A_\m$),  
\bea
0 & =&  \delta_\LL \left\{ \G^\n_{ab} A_{\n,\m} + v_\m + B^{(\G^\n_{ab} A_\n + v)}_\m \right\}
 - \delta_{(\G^\n_{ab} A_\n + v)} B^\LL_\m, \cr
& =& \delta_{f(\LL,\G^\n_{ab} A_\n + v)} A_\m + 
 \delta_\LL (\G^\n_{ab}A_{\n,\m} + v_\m) + B^{(\delta_\LL (\G^\n_{ab}A_\n+v))}_\m.
\eea
Collecting the undifferentiated $\LL$ terms at $O(\T^k)$ gives 
$$0 = (b^k_{0})_{\n, \m}.$$
This in turn implies the vanishing of $(b^k_{0})_\n$ and consequently of $c^k_0$.  
By field redefinitions, we can therefore transform $B$ and $C$ into their expected
forms.

\section{Constraints in Commutative Variables}

As argued in \cite{Seiberg:1999vs}, the non-commutative action should be expressible in terms of the
commutative variables. We are left with the following puzzle: 
if the deformation of the gauge symmetry is so strongly constrained
by supersymmetry, what is the corresponding constraint in commutative variables? The Lagrangian 
in commutative variables seems to take a fairly generic form with an 
infinite number of higher derivative terms that depend on $F$~\cite{Ganor:2000my} (see, 
also,~\cite{Rey:2000hh}), 
$$ L = F^2 + \T F^3 + \ldots. $$
To address this question, we begin by assuming that
\begin{equation}\label{changevar} \hA = \hA (A), \quad \hF = \hF (A, \F), \end{equation}
where the hatted variables are non-commutative variables, while the unhatted variables 
are commutative. The map between non-commutative and commutative variables can be constructed
using the techniques of section \ref{detbc} once the map between $\widehat{\LL}$ and
$\LL$ is known. In particular, once one has transformed the gauge parameters, 
equation \C{nconstraints} holds for $B^k_{ex}=0.$ We can then repeat the analysis of
that section to inductively construct the commutative variables. At least at the classical
level, that construction gives
a map from non-commutative to commutative fermions which 
is linear in the fermions (although non-linear in the gauge bosons). 

Using these observations, the change of variables given in~\cite{Seiberg:1999vs} can 
easily be extended to include fermions. To $O(\T)$, the change of variables is
given by:
\bea\label{explchange} 
 & \hA_\m  & = A_\m + {1\over 2} \T^{ij} A_i \left( \p_j
A_\m + F_{j\m} \right) + O(\T^2), \cr
& \widehat{F}_{\m\n} & = F_{\m\n} + \T^{ij} \left( F_{\m i} F_{j \n} +
A_{i} \p_j F_{\m\n}\right)+ O(\T^2), \cr
 & \hF & = \F + \T^{ij} A_i \p_j \F + O(\T^2), \cr
& \widehat{\LL} & = \LL + {1\over 2} \T^{ij} A_i \p_j \LL+ O(\T^2).
\eea
{}From the supersymmetry transformations, 
\bea \label{noncommsusy} & \dd_\e \hA_{\mu} & ={ i \over 2}\,  \ce \, \G_{\mu} \hF, \cr 
& \dd_\e \hF  & = - {1 \over 4} \widehat{F}_{\mu \nu} \G^{\mu \nu} \e,  \eea
we can derive transformations for the commutative variables. We learn about $N$ from the $\hA$ variation,  
$$ \dd_\e \hA_\m = 
\dd_{A_\n} \hA_\m \, \left( \dd_\e A_\n \right)
=   \dd_{A_\n} \hA_\m \left(   \ce \, \left(  { i \over 2}\G_{\n} + N_\n \right) \F \right)
 = { i \over 2}\,  \ce \, \G_{\mu} \hF,   $$
where we have used our usual notation,
$$ \dd_\e A_{\mu}  =   \ce \, \left(  { i \over 2}\G_{\mu} + N_\m \right) \F. $$
This gives the relation, 
\bea
& (\G^0 N_\m\F) + { i \over 2} \T^{ij} (\G^0\G_i \F) \left( \p_j
A_\m - { 1 \over 2}\p_\m A_j \right) +  { i \over 2} \T^{ij} A_i \left(\G^0\G_\m \p_j \F
-{1\over 2} \G^0 \G_j \p_\m\F \right)  & \cr & = { i \over 2} \G^0 \G_\m  \T^{ij} A_i \p_j \F. & \cr
&  \Rightarrow \quad N_\m \F =  { i \over 4} \T^{ij} A_i \left( \G^0 \G_j \p_\m\F \right) - 
{ i \over 2} \T^{ij} (\G^0\G_i \F) \left( \p_j A_\m - { 1 \over 2}\p_\m A_j \right).
\eea
This teaches us that $N$ is bosonic, and this must be the case to all orders in $\T$ because
the transformation between $\hF$ and $\F$ is linear in the fermion field. Note that, curiously, $N$
is {\it not} gauge invariant.  

On the other hand, we learn about $M$ from the $\hF$ variation which gives the relation, 
$$  \left(M_{ba} -{1\over 4} \T^{ij} A_i \p_j F_{\m\n}\G^{\m\n}_{ba} \right) 
+ {i\over 2} \T^{ij} (\G^0\G_i \F)_a \p_j \F_b = -{1\over 4} \left\{ \T^{ij} \left( F_{\m i} F_{j \n} +
A_{i} \p_j F_{\m\n}\right) \right\} \G^{\m\n}_{ba}.$$
This leads to the expression, 
\begin{equation}\label{newM} M_{ba} =  -{1\over 4} \left\{ \T^{ij} F_{\m i} F_{j \n}\right\} \G^{\m\n}_{ba}
- {i\over 2} \T^{ij} (\G^0\G_i \F)_a \p_j \F_b. \end{equation}
The special features of this supersymmetry algebra are now visible. 
The fact that $N$ is bosonic and that the transformation 
between $\hF$ and $\F$ is linear in $\F$ implies that $M$ can never be
more than quadratic in fermions. Such an $M$ is morally of the form we should expect for a non-linear 
sigma model. We can view $M$ as roughly determining a connection. However, the action in commutative
variables only contains quadratic couplings in the fermions, again because $\hF$ is linear in $\F$. 
This suggests that the ``curvature'' of the connection (which would be reflected in a four fermion coupling)
is zero. This is quite different from what we 
might have expected from a theory with an infinite number of higher derivative interactions. For
example, supersymmetric DBI has many couplings involving large strings of 
fermions~\cite{Aganagic:1997nn}.

This algebra also violates our constraint \C{bigdeal} that the gauge charges be gauge-invariant.
Rather,
\bea \label{bizarre}
[\dd_\LL, \dd_\e] A_\m &=& \dd_\LL (\e^T \G^0 N_\m \F), \cr
&=& \p_\m ( -{i\over 4} \T^{ij} \p_i \LL \, \e^T \G_j \F ), \cr
& = & \dd_{\LL'} A_\m,
\eea
where $\LL'$ is a field-dependent gauge transformation. The supersymmetry algebra now takes the form,
\begin{equation}
[\dd_\LL, \dd_\e] = \dd_{\LL'}. 
\end{equation}
Note that $M$ is gauge-invariant. We require that
the fermion equation of motion be gauge-invariant. However, the fermion equation of motion is needed
{}for closure of the commutator,  
$$ [\dd_\e, \dd_{\e'}] \F,$$
on Poincar\'e. 
The only place that a problem might arise at $O(\T)$ is from 
$$\dd_\e F_{\m\n} \G^{\m\n} {\e'} - \dd_{\e'} F_{\m\n} \G^{\m\n} {\e}.$$ 
What appears at $O(\T)$ is the combination, 
$$ \p_\m N_\n \F - \p_\n N_\m \F. $$
Using \C{bizarre}, we see that this is indeed gauge-invariant. This guarantees that the fermion
equation of motion is gauge-invariant, as we expect. It seems possible that there is a connection
between the non-gauge invariance of $\dd_\e$, and the observation of~\cite{Gross:2000ph} that
translations in the non-commutative directions are equivalent to gauge transformations. 
Whether the special structure of this supersymmetry algebra extends to the $(2,0)$ 
theory of a six-dimensional self-dual tensor multiplet remains to be seen~\cite{toappear}.

\section{Acknowledgements}
It is our pleasure to thank N. Itzhaki,  
H. Liu, and, N. Seiberg for helpful discussions. 
The work of S.~P. is supported in part by 
NSF Grant No. PHY-0071512,  while
the work of S.~S. is supported in
part by NSF CAREER Grant No. PHY-0094328, and 
by the Alfred P. Sloan Foundation. The work of M.~S. 
is supported in part by NSF Grant No. DMS-9870161.

%\newpage
%\bibliographystyle{amsunsrt-es}
%\bibliography{myrefs}

\end{document}